\documentstyle[prl,aps,multicol,epsf]{revtex}
\begin{document}                                                        
\draft                     
\title{Kondo--lattice-like effects of hydrogen in transition metals}

\author{R. Eder, H. F. Pen, and G. A. Sawatzky}
\address{Department of Solid State Physics, 
Materials Science Centre,\\
University of Groningen,
9747 AG Groningen, The Netherlands}
\date{\today}
\maketitle

\begin{abstract} 

We discuss the possibility of a Kondo like effect associated with H in metals
resulting from the strong dependence of the H $1s$ orbital radius on the
occupation number.  We demonstrate that such a strong breathing property of the
orbital radius, which translates directly into a strong occupation dependent
hopping, results in the formation of local singlet-like bound states involving
one electron on H and one on the surrounding metal orbitals.  We also show that
already at a mean field level an occupation dependent hopping integral leads to
a substantial potential energy correction on hydrogen, and that the failure of
band structure methods to incorporate this correction is responsible for the
incorrect prediction of a metallic ground state for the YH$_3$ switchable mirror
compounds.

\end{abstract} 
\begin{multicols}{2}
\section{Introduction}

The recent discovery\cite{Huiberts96} of the so called switchable mirror
compounds based on YH$_{3-x}$ has renewed the interest in the electronic
structure of transition metal and rare earth hydrides.  These compounds undergo
a metal to insulator transition as $x$ changes from one to zero with the
accompanying change from high optical reflectivity to optical transparency for
photon energies below about 2 eV.  LDA band structure calculations fail to
reproduce the semiconducting gap for $x=3$ in the widely accepted HoD$_3$
structure or in the simpler cubic LaH$_3$ structure.  Although a gap can be
obtained for more complicated distorted structures\cite{Kelly97}, it is still
much too small; and although such distortions are not excluded for YH$_3$, they
do not appear to occur for LaH$_3$, which has similar properties.

The failure to produce large enough band gaps in semiconductors by LDA is a
well-known shortcoming which here however seems to take quite dramatic forms:
The valence and conduction bands in LDA overlap by about 1 eV, so a total
relative shift of about 3 eV is required to match the experimental value.  Such
dramatic discrepancies are reminiscent of the strongly correlated systems like
the transition metal and rare earth oxides and point perhaps to the importance
of correlation effects.  Since however the H $1s$ orbitals are rather extended
as compared to the $3d$'s of the transition metals, especially for the negative
ion, they are expected to form rather broad bands, and the on--site Coulomb
interactions are strongly screened.  Therefore the origin of the correlation
effects may be quite different.  In this paper we address this problem and come
to the new suggestion that the correlation effects are a consequence of the
large change in the H~$1s$ orbital radius upon orbital occupation.  This
``breathing'' property of the hydrogen ion is shown to introduce a new term in
the mean field treatment of the electronic structure of hydrides.  This term
results in an opening up of the band gap in a quite natural way, with the
retention of large band widths and nearly one particle behaviour of the excited
states.  Using a model Hamiltonian to demonstrate this behaviour we also show
that for a range of parameter values the system behaves like a Kondo lattice
insulator similar to that suggested by Ng {\it et al}\cite{Ng97}.

\section{The breathing hydrogen atom}

\begin{figure}
\epsfxsize=2.0truein 
\centerline{\epsffile[68 302 558 744]{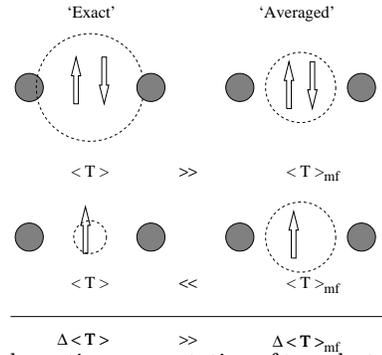}}
\narrowtext 
\caption[]{Schematic representation of two electron and single
electron wave function with (left column) and without (right column) taking into
account the expansion of the Hydrogen wave function.}  \label{mf} 
\end{figure}

As is well known and referred to in most general chemistry text books, the so
called effective radius of hydrogen is extremely strongly dependent on the
charge state.  The crystal radius of neutral H is 0.26~\AA , whereas that of the
negative ion H$^-$ is 1.54~\AA\onlinecite{McQuarrie84}.  The values of the
average $1s$ orbital radius $\sqrt<r^2>$, as obtained from free ion
Hartree--Fock calculations, are 0.8 \AA~ for H and 1.72 \AA~ for
H$^-$\onlinecite{Jonkman00}.  This very large change is not unexpected since in
H, with its low nuclear charge of one, the screening of the nuclear Coulomb
potential by a second $s$ electron is very important and has a dramatic effect
on the orbital radius.  This large effect causes the effective hopping integrals
or hybridizations with surrounding ions to be strongly different for the
fluctuations involving H to H$^+$ as compared to those involving H to H$^-$, as
pictured in Fig.\ \ref{mf}.
\noindent If these instantaneous changes in the hopping integrals are larger
than or comparable to other energy scales, like the orbital energy splittings,
they must be treated explicitly.  They can not be treated in a mean field like
way (as is done in band theory), taking an average orbital radius corresponding
to the average occupation as determined from self--consistent calculations.

\begin{figure}
\epsfxsize=3.0 truein
\centerline{\epsffile[53 490 519 674]{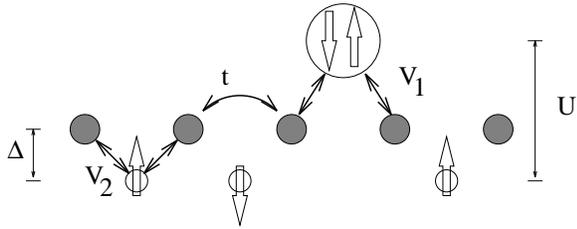}}
\narrowtext
\caption[]{Schematic representation of the 1D
model used in the exact diagonalization.}
\label{model}
\end{figure}

\begin{figure}[thb]
\epsfxsize=2.75truein
\centerline{\epsffile[43 189 474 706]{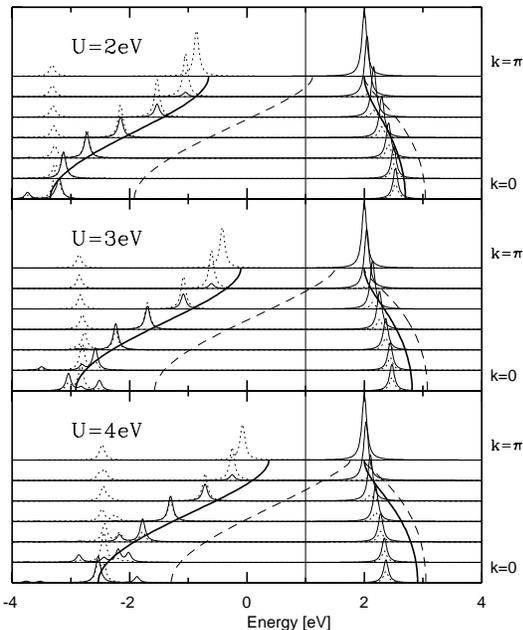}}
\narrowtext
\caption[]{Single particle spectral function of the 1D
model with six unit cells. The part to the right (left)
of the thin vertical line correspond to electron addition (removal)
from the half-filled ground state (i.e., two electrons per unit cell).
The full line corresponds to electron
removal or addition on the ``metal'' sites, the dashed line to
hydrogen. Parameter values are
$V_1=2$~eV, $V_2=0.2$~eV, $t=0.5$~eV, $\Delta=1$~eV (see Fig. \ref{model}).
The full (dashed) dispersion curve gives the mean-field bands
calculated with (without) the correction $V_T=\alpha \langle T\rangle$.}
\label{spectra}
\end{figure}

To model the effect of the ``breathing'' hydrogen we introduce an occupation
dependent hopping integral (following Hirsch\cite{Hirsch93a})
between H and its nearest neighbours, in addition to the usual on-site Coulomb
interaction of the Hubbard\cite{Hubbard63} or Anderson impurity\cite{Anderson61}
models. We consider the Hamiltonian of the form:
\begin{eqnarray}
H &=& 
- \Delta \sum_\sigma h_{\sigma}^\dagger h_{\sigma}
+ U h_{\uparrow}^\dagger h_{\uparrow}
h_{\downarrow}^\dagger h_{\downarrow}
\nonumber \\
&-&  \sum_\sigma \;[\;(\;
 V_1 l_{\sigma}^\dagger 
h_{\sigma} h_{\bar{\sigma}}^\dagger h_{\bar{\sigma}}
+ V_2 l_{\sigma}^\dagger
h_{\sigma} h_{\bar{\sigma}} h_{\bar{\sigma}}^\dagger\;)
+ H.c.\;],
\end{eqnarray}

which describes the hybridization of a single hydrogen atom with a single ligand
orbital.  Here $h_{\sigma}^\dagger$ ($l_{\sigma}^\dagger$) creates an electron
on the hydrogen (ligand), there is a charge transfer energy $\Delta$ between
hydrogen and ligand (we assume $\Delta>0$) and a Coulomb repulsion $U$ between
electrons on hydrogen.  $V_1$ and $V_2$ are the occupation dependent
hydrogen--ligand hopping integrals.  We discuss the first ionization energy
$E_-$ for photoemission on a single cell, consisting of a hydrogen atom and a
ligand orbital, filled with two electrons.  The minimum ionization energy $E_-$,
together with the minimal electron affinity $E_+$, determine the excitation gap
$E_{gap} = E_- - E_+$.  An underestimation of the magnitude of $E_-$ thus may
lead to a too small gap energy, as seems to be the case in LDA calculation for
YH$_3$.\\ For simplicity we take $V_2 =0$, so that the ground state of a single
electron just corresponds to the electron being trapped in the collapsed
hydrogen orbital, and has energy $-\Delta$.  A straightforward calculation then
gives the corresponding ionization energy
\[
E = \frac{U-\Delta}{2} - \sqrt{ (\frac{U-\Delta}{2})^2 + 2V_1^2 }.
\]
Taking for simplicity $U=\Delta$ we obtain $E = -\sqrt{2} V_1$.  The ionization
energy thus is predominantly due to the loss of kinetic energy, because the
single electron in the final state cannot escape from the collapsed hydrogen
orbital, so that the large gain in kinetic energy, which was possible for two
electrons, is no longer possible.\\ On the other hand, constructing a
single-particle Hamiltonian with an averaged hopping integral $V_{MF} \approx
\langle h_\sigma^\dagger h_\sigma \rangle V_1$ and an ``effective'' on-site
energy $\Delta_{MF}$ (as it is done in an LDA calculation), the excitation
energy would be simply the energy of the occupied mean-field orbital, i.e.,
\[ 
E_{MF} = \frac{\Delta_{MF}}{2}
- \sqrt{ (\frac{\Delta_{MF}}{2})^2 + 2 V_{MF}^2 }.
\]
If the occupation of hydrogen $1s$ is significantly smaller
than one per spin state, this way of calculation will miss a large part of the
kinetic energy contribution to the excitation energy,
unless the ``effective'' on-site energy is corrected to take this
effect into account.\\
In a mean-field treatment
of this Hamiltonian, it will become apparent that
the occupation dependent hopping gives rise to very peculiar
physics. Breaking down the conditional hopping terms
into quadratic terms we obtain:
$l_{\sigma}^\dagger h_{\sigma} h_{\bar{\sigma}}^\dagger h_{\bar{\sigma}}
\rightarrow
l_{\sigma}^\dagger h_{\sigma} \langle
h_{\bar{\sigma}}^\dagger h_{\bar{\sigma}} \rangle +
\langle l_{\sigma}^\dagger h_{\sigma}\rangle
 h_{\bar{\sigma}}^\dagger h_{\bar{\sigma}} $.
The first of these terms corresponds to weighting the
``large'' hybridization integral by the occupation of the
hydrogen orbital, which is what one might have expected;
the second term, however, is a correction to the on-site energy
of hydrogen by a part of the kinetic energy. All in all we obtain:
\begin{eqnarray}
H_{MF} &=& 
\sum_\sigma \;[\; - \Delta_{MF} h_{\sigma}^\dagger h_{\sigma}
+ \; (\;V_{MF} l_{\sigma}^\dagger
h_{\sigma} + H.c.\;)\;],
\nonumber \\
V_{MF} &=& V_1\; n_H + V_2\; (1-n_H)
\nonumber \\
-\Delta_{MF} &=& -\Delta + U\; n_H + \alpha \langle T \rangle
\nonumber \\
\alpha &=& \frac{V_1 - V_2}{V_{MF}}
\label{hmf}
\end{eqnarray}
where
$n_h = \langle h_{\sigma}^\dagger h_{\sigma} \rangle$ and
$\langle T \rangle$ is the energy of hybridization between
hydrogen and ligand. We thus find the surprising result
that in this approximation the expectation value of the
kinetic energy
$\langle T \rangle$ enters as an additional ``potential''
on the hydrogen sites, and is in fact even enhanced by
the factor $\alpha$. In the limit
$V_1 \gg V_2$ we find $\alpha \rightarrow n_H^{-1} > 1$, so that
the correction to the on-site potential of hydrogen becomes
$(V_1/V_{MF}) \langle T \rangle$, i.e., the kinetic energy
for mixing with the hydrogen site, but calculated with the
hopping integral for the ``expanded'' atom. This is clearly a
huge energy, but it has a very clear physical interpretation:
for two electrons in the cell, the hydrogen atom will oscillate
between $H^0$ and $H^-$, so as to take maximum advantage of the
expansion of the wave function, and the hybridization energy will
be large. Removing one electron, the
remaining electron will essentially be trapped in the ``collapsed''
hydrogen orbital, and there is practially no more hybridization energy.
In the mean-field wave function both electrons are in the lower
molecular orbital, which (due to its strongly negative
effective on-site energy resulting from Eq.~\ref{hmf}) has predominant hydrogen character.
The ionization energy, which by Koopmans theorem should be
given by the mean-field eigenvalue, then contains
almost the entire kinetic energy of the two-electron state,
because this kinetic energy has been put into the on-site potential
of the hydrogen atom. \\
To make this more quantitative,
we have performed exact diagonalization calculations
for a 1D chain of a model with occupation dependent
hybridization between hydrogen and ligand.
A schematical representation of the model is given in
Fig. \ref{model}. 
\noindent 
Computer memory and CPU time limitations 
prohibit to diagonalize chains of more than six unit cells
of this model, at least for the most interesting densities
near two electrons per unit cell. For the given parameters, the
ground state of this system corresponds closely to a state
with two electrons in a local singlet state in each unit cell,
with one electron primarily on H and the other primarily on 
the neighbouring ligands. Such a state is reminiscent of a
Kondo-lattice insulator ground state.
Figure \ref{spectra} shows the
$\bbox{k}$-resolved electron addition and removal spectra for
the six-site chain at half-filling. To get a feeling for the 
dispersions of the different bands, we have combined spectra
calculated with both periodic and antiperiodic boundary
conditions. While there is no rigorous proof for this,
inspection of Fig. \ref{spectra} shows that in this way one obtains
remarkably smooth ``band structures''.
\noindent
Unlike spectra for, e.g., the Hubbard model at half-filling,
the calculated spectra are surprisingly ``coherent'', with almost
all of the spectral weight being concentrated in just three
well defined ``bands''. In inverse photoemission there is
a band of predominant metal character, with little dispersion.
In photoemission, there is a strongly dispersive band of
mixed hydrogen-metal character, and a dispersionless
low intensity band of practically pure hydrogen character.
More detailed analysis shows, that the dispersionless band
corresponds to $H^+$ final states (i.e., it is a kind of
``lower Hubbard band'') whereas the dipersive band
corresponds to $H^0$-like final states.
Next, Fig. \ref{spectra} shows the spectral function
for different values of $U$. For comparison, the bands obtained
by a mean-field solution of the model are also shown.
Thereby the calculation has been done both with and without
the kinetic energy correction $V_{T}=\alpha \langle T\rangle$
to the hydrogen on-site
potential. Quite obviously the calculation with $V_T$
reproduces the exact band structure very well, whereas the
bands without $V_T$ while giving roughly the correct dispersion
substantially underestimate the gap size. As explained above,
 we believe that LDA misses the kinetic energy correction $V_T$,
so that the LDA band structure rather corresponds to the bands
without $V_T$. In a phenomenological way, this suggests
a kind of ``scissors operator'' approach to obtain 
the ``correct'' band structure of YH$_3$ from the LDA result.

\section{Application to YH$_3$}

\begin{figure}
\epsfxsize=2.75truein
\centerline{\epsffile[55 332 476 706]{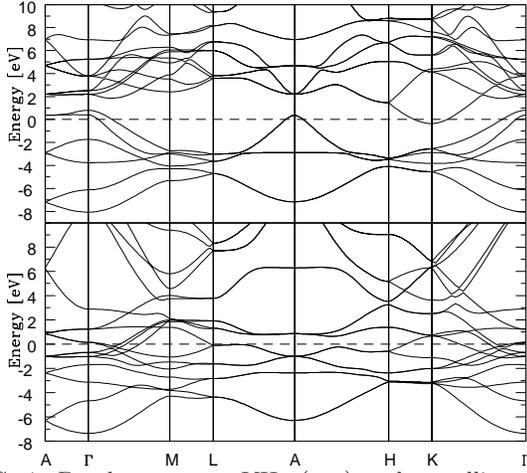}}
\narrowtext
\caption[]{Band structures YH$_3$ (top)
and  metallic yttrium with the 
lattice constants of YH$_3$ (bottom). The Fermi energy
is marked by the dashed line.}
\label{yband}
\end{figure}

We now use the above ideas for the case of YH$_3$ and attempt to obtain
reasonable parameters and subsequent estimates of the band gap.  
The three $5d$ electrons of Y will in the above scenario all
be bound by the three H atoms per Y; this would again result in an 
insulating ground state. First, we obtain
good estimates for the average hopping integrals and on--site energies, using a
tight binding fit to an LDA band structure calculated with the LMTO
method\cite{Andersen75}.  The upper panel of Fig.\ \ref{yband} shows the band
structure calculated for YH$_3$ in the LaH$_3$ structure, which is practically
identical to previous published results\cite{Wang93,Dekker93}.  In the lower
panel we show the result of a calculation for Y metal with the lattice constants
of YH$_3$, in order to establish the contribution of the H $1s$ orbitals.

To begin with, the lowest two bands which have
predominant $5sp$ character, are
nearly identical in both solids. Obviously, these free--electron-like
states are not significantly affected by the insertion
of hydrogen into the interstitial sites.
In yttrium metal, the next group of states are the Y~$4d$ bands,
which form a dense continuum with a width of $\sim 6$~eV.
The Fermi energy cuts into the lower part of this complex of
$d$ bands. In YH$_3$ the situation is very different:
the lowest Y~$4d$ band (which is still quite reminiscent
of the third-lowest band in metallic yttrium)  is split off
from the remaining $d$ bands by
$\sim 4$~eV, and in the resulting gap bands of
predominantly hydrogen character are inserted.
Although these hydrogen bands do have an appreciable width,
they barely overlap with the lowermost of the following
$d$ bands. In fact, the band structure of YH$_3$ already
shows a very clear ``gap'' between the
top of the hydrogen like valence band and the Y~$4d$-like
conduction band throughout the entire Brillouin
zone -- YH$_3$ thus is already ``almost'' a semiconductor. The Fermi 
energy cuts into the top of the hydrogen-like valence band and the
bottom of the $d$-like conduction band,
so that LDA predicts YH$_3$ to be a semimetal.
The shift of the lowermost $d$ to considerably
higher binding energy upon insertion of hydrogen,
which is predicted by LDA, is in qualitative agreement with
the photoemission data of Fujimori and Schlapbach\cite{Fujimori84}:
For metallic yttrium, these authors found a high intensity
structure at binding energies $\leq 2$~eV, which probably
corresponds to the occupied part of the Y~$4d$ bands. For
YH$_3$ a similar structure occurs at a binding energy of
6~eV, indicating the shift of the $d$ band away from
$E_F$. It should be noted, however, that the
experimental shift is larger by $\approx 2$~eV than that
predicted by LDA. The picture thus is quite reminiscent
of the well-known band-gap problem in semiconductors,
where LDA fails to give correct values
for the semiconducting gaps.

\begin{figure}
\epsfxsize=2.75truein
\centerline{\epsffile[29 394 478 706]{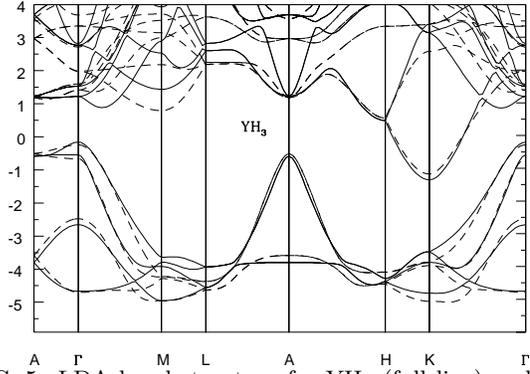}}
\narrowtext
\caption[]{LDA band structure for YH$_3$ (full line)
and tight binding fit (dashed line). Energies are in eV.}
\label{tbfit}
\end{figure}

\begin{figure}
\epsfxsize=1.5truein
\centerline{\epsffile[60 444 300 746]{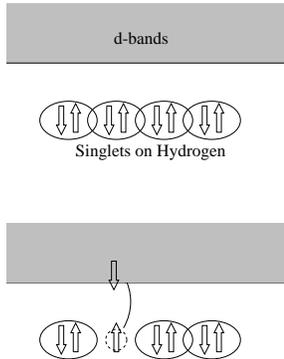}}
\narrowtext
\caption[]{Schematic representation of the stabilization
energy for the local two-electron bound state.}
\label{stab}
\end{figure}

\begin{figure}
\epsfxsize=2.5truein
\centerline{\epsffile[18 376 412 706]{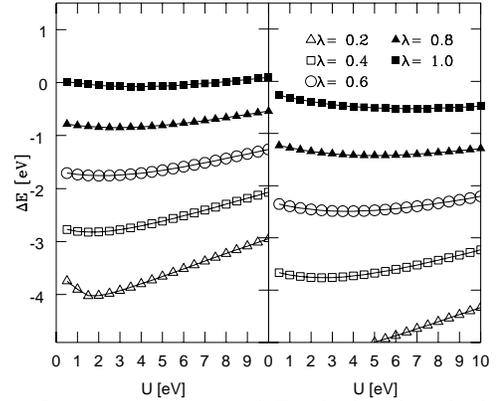}}
\narrowtext
\caption[]{Ionization energies $\Delta E$ calculated with the
impurity model for the metal plane hydrogen (left panel) and the
tetragonal hydrogen (right panel). Note that the
other tight binding parameters depend on the values for $U$ and
$\lambda$ through Equations
(\protect\ref{TBpar}) and (\protect\ref{BareOnsite}).}
\label{eb}
\end{figure}

To extract additional information, we performed a 
 tight-binding fit to the LDA band structure.
It turned out that by using hybridization integrals
only between nearest neighbours a surprisingly good fit
of the first few valence and conduction bands
could be obtained, as shown in Fig. \ref{tbfit}.
The tight binding fit was
obtained using the following Hamiltonian in a mean field way, as discussed below:
\begin{eqnarray}
H &=&
\sum_{\bbox{k},\nu,\sigma} \epsilon_\nu(\bbox{k})
d_{\bbox{k},\nu,\sigma}^\dagger d_{\bbox{k},\nu,\sigma}
+ \sum_{j,\sigma} \tilde{\epsilon}_j
h_{j,\sigma}^\dagger h_{j,\sigma}
\nonumber \\
&+&
\sum_{i,j,\sigma} [\;(\;
V_{(i,\nu),j}^{(1)} d_{i,\nu,\sigma}^\dagger h_{j,\sigma}
h_{j,\bar{\sigma}}^\dagger h_{j,\bar{\sigma}}  +
\nonumber \\
&\;&\;\;\;\;\;\;\;\;\;\;\;\;\;
V_{(i,\nu),j}^{(2)} d_{i,\nu,\sigma}^\dagger h_{j,\sigma}
h_{j,\bar{\sigma}} h_{j,\bar{\sigma}}^\dagger\;) + H.c.\;]
\nonumber \\
&+&
U \sum_j h_{j,\uparrow}^\dagger h_{j,\uparrow}
h_{j,\downarrow}^\dagger h_{j,\downarrow}.
\label{full}
\end{eqnarray} 
Here $\epsilon_\nu(\bbox{k})$ denotes the Y~$4d$ bands, which 
we obtained from the tight-binding fit.\\
In a mean-field treatment (\ref{full}) would turn into
\begin{eqnarray}
H &=&
\sum_{\bbox{k},\sigma} \epsilon_\nu(\bbox{k})
d_{\bbox{k},\nu,\sigma}^\dagger d_{\bbox{k},\nu,\sigma}
+ \sum_{j,\sigma} \Delta_j
h_{j,\sigma}^\dagger h_{j,\sigma}
\nonumber \\
&+&
\sum_{i,j,\sigma} [\;(\;
V_{(i,\nu),j}^{(1)} d_{i,\nu,\sigma}^\dagger h_{j,\sigma}
n_j +
\nonumber \\
&\;&\;\;\;\;\;\;\;\;\;\;\;\;\;
V_{(i,\nu),j}^{(2)} d_{i,\nu,\sigma}^\dagger h_{j,\sigma}
(1-n_j) \;) + H.c.\;],
\end{eqnarray}  
with $ n_j = \langle h_{j,\uparrow}^\dagger h_{j,\uparrow} \rangle
= \langle h_{j,\downarrow}^\dagger h_{j,\downarrow} \rangle$
and $\Delta_j = \tilde{\epsilon}_j + U n_j$.
We now introduce the parameter
$\lambda$, which we assume independent of $j$, 
as $\lambda= V_{(i,\nu),j}^{(1)}/V_{(i,\nu),j}^{(2)}$,
i.e., the ratio of hopping integrals for the collapsed and
expanded hybdrogen atom. We then estimate the change as
\begin{equation}
V_{(i,\nu),j}^{TB} \approx ( n_j + \lambda (1-n_j)) V_{(i,\nu),j}^{(1)}, \label{TBpar}
\end{equation}
where $V_{(i,\nu),j}^{TB}$ is the hybridization integral
extracted from the tight-binding fit.
Since $n_j$ can be obtained from the tight-binding
calculation as well, we can thus, for given $\lambda$, obtain an
estimate of $V_{(i,\nu),j}^{(1)}$.
Next, we estimate the ``bare'' on-site energies $\epsilon_j$
of the hydrogen atoms from those of the tight-binding fit,
$\Delta_j^{TB}$, as follows:
\begin{equation}
\epsilon_j  = \Delta_j^{TB} - n_j U. \label{BareOnsite}
\end{equation}
This introduces another unknown parameter: the onsite
Coulomb repulsion $U$ between two electrons on the
hydrogen site. While an ab-initio calculation of $U$ and
$\lambda$ would be highly desirable, this is outside the
range of techniques available to us.
We therefore will treat these quantities as
implicit parameters, and consider the variation of
possible results when $U$ and $\lambda$ are varied within
``reasonable bounds''.\\
Using the parameters estimated in this way, we now proceed to
an impurity--like calculation to determine the stabilization energy of Kondo-like
local singlets, formed on a single hydrogen atom
in the lattice of Y~$4d$ orbitals.
In the first step, we calculate the ground state energy $E_0^{(2)}$ of
a two-electron bound state from the ansatz
\[
|\Psi^{(2)}\rangle = 
\;[\;\alpha h_\uparrow^\dagger h_\downarrow^\dagger +
\frac{1}{\sqrt{2}} \sum_{\nu, \bbox{k}}
\beta_\nu ( \;
d_{\bbox{k},\nu,\uparrow}^\dagger
h_\downarrow^\dagger +
h_\uparrow^\dagger
d_{\bbox{k},\nu,\downarrow}^\dagger\;)\;] |vac\rangle.
\]
Here summation over $\bbox{k}$ and $\nu$ refers to
the Y~$4d$ bands. The dispersion of these bands
and the hybridization matrix elements between
the band states and the hydrogen atom are calculated using the
parameters from the tight binding fit,
whereby the hydrogen-yttrium hybridizations depend
on $\lambda$ and the hydrogen on-site energy on $U$.
Then, we want to know the stability of this state against decay
into a state with a single electron remaining in the
hydrogen atom, and the second electron being
in a free yttrium $d$-like state. The energy $E_0^{(1)}$ of the 
single electron in the hydrogen is calculated from the ansatz
\[
|\Psi^{(1)}\rangle = 
\;[\;\alpha' h_\uparrow^\dagger +
\sum_{\nu, \bbox{k}}
\beta_\nu' 
d_{\bbox{k},\nu,\uparrow}^\dagger \;] |vac\rangle,
\]
and for the energy of the $d$-like electron we simply choose the
lower bound of the $d$-band complex, $E_{edge}$.
We then form the difference 
$\Delta E= E_0^{(2)} - (E_0^{(1)} + E_{edge})$ (see Fig. \ref{stab}),
\noindent
which
obviously determines the stability of the two-electron state
against decay. This energy will be a function of the
unknown parameters $U$ and $\lambda$.
The result then is shown in Fig. \ref{eb}.\\

One can see that for ``reasonable'' values of $U$ and
not very extreme values of $\lambda$ the two-electron
bound state attains a stabilization energy of several electronvolts.
Drawing an analogy with the situation in cuprate superconductors,
where the Zhang-Rice singlet has a stabilization energy
of approximately 1~eV, it seems quite reasonable to
adopt the picture of local bound states. Then, for
YH$_3$ one may expect that these bound states form a split-off band,
with the Fermi energy lying in the gap between these states
and the bottom of the $4d$ band; the physics
is similar to our exact result on the 1D cluster.
\noindent 

\section{Doping dependence of the electronic structure}

\begin{figure}
\epsfxsize=2.5truein
\centerline{\epsffile[85 144 442 496]{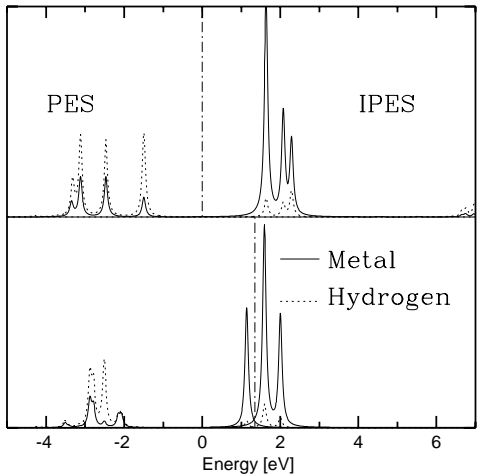}}
\narrowtext
\caption[]{Single particle $\bbox{k}$-integrated spectral function for a three unit cell
cluster of the 1D model with open boundary conditions.
The spectra are calculated at ``half-filling'' (upper part)
and with one charge neutral hydrogen atom removed
from the central cell (lower part). The parts of the spectra to the right
(left) of the vertical dashed-dotted line correspond to electron
addition (removal). Parameter values are $\Delta=1$~eV, $U=2$~eV,
$t=0.5$~eV, $V_1=2$~eV, $V_2=0.2$~eV.}
\label{doped}
\end{figure}

To simulate the physics of what happens as one removes hydrogen from YH$_3$, we
resort to an exact calculation of a small cluster as done above, but we now
remove one hydrogen site and one electron.  In Fig.\ \ref{doped}
we show the spectral function for electron removal and electron addition
before and after removal of the hydrogen.  In the top panel, one can clearly
recognize the large gap between the hydrogen like valence band and the metal
like conduction band.  Upon hydrogen removal, as shown in the bottom panel, the
Fermi energy jumps into the metal band which implies that hydrogen behaves like
an H$^-$ ion in that it binds two electrons. This is consistent with the above 
discussion and also suggested by Ng {\it et al}\cite{Ng97}.  

In our previous discussion we came to the conclusion that H actually binds two
electrons:  one localized on H, and the other on the nearest neighbour metal atoms.
With the removal of a H atom, which takes only one electron with it, another
``lonely'' electron is left behind, which must then be in the conduction band.
Removal of H from the trihydride insulator should then transfer spectral weight for
electron removal from the top of the valence band to the bottom of the conduction
band, some 2 eV higher in energy.  This kind of behaviour upon doping is very
similar to that predicted for\cite{Eskes91} and observed\cite{Chen91} in the high
$T_c$ cuprates.  Consistent with this are the observations by Peterman {\em et
al.}\cite{Peterman81}, who found that in hydrogen depleted trihydrides the Fermi
energy falls into a ``band'' with very weak spectral weight, which grows upon
further depletion.

\section{Conclusions}

We have shown that hydrogen is an extreme example of an atom with a large orbital
occupation dependence of the orbital radius, leading to large occupation dependent
hopping integrals in hydrides.  We argue that explicit inclusion of such terms in
the Hamiltonian results in a ``scissors operator'' like separation of the valence
and conduction bands and, consequently, the opening of a substantial gap.  We argue
that the insulating character of YH$_3$ can be understood in this way.  Using
reasonable parameters obtained from tight binding fits to the band structure, we
find that the ground state of YH$_3$ corresponds closely to that of a Kondo
insulator, with each H binding two electrons in a singlet state.

\noindent 

\end{multicols}
\end{document}